\documentclass[twocolumn,showpacs,preprintnumbers]{revtex4}%
\usepackage{amsfonts}
\usepackage{amsmath,epsfig}
\usepackage{graphicx}
\usepackage{dcolumn}
\usepackage{bm}
\usepackage{amssymb}
\usepackage{amsmath}%
\begin{document}
\title{Oscillations in the decay law: A possible quantum mechanical explanation of
the anomaly in the experiment at the GSI facility}
\author{Francesco Giacosa$^{\text{(a)}}$and Giuseppe Pagliara$^{\text{(b)}}$}

\begin{abstract}
We study the deviations from the usual exponential decay law for quantum
mechanical systems. We show that simple and physically motivated deviations
from the Breit-Wigner energy distribution of the unstable state are sufficient
to generate peculiar deviations from the exponential decay law. Denoting with
$p(t)$ the survival probability, its derivative $h(t)$ shows typically an
oscillating behavior on top of the usual exponential function. We argue that
this can be a viable explanation of the observed experimental results at GSI
Darmstadt, where the function $h(t)$ has been experimentally measured for
electron capture decays of Hydrogen-like ions. Moreover, if our interpretation
is correct, we predict that by measuring $h(t)$ at times close to the initial
one, the number of decays per second rapidly drops to zero.

\end{abstract}

\pacs{03.65.-w,03.65.Xp,23.40.-s}
\keywords{Non-exponential decay, GSI anomaly}\maketitle

\address{$^{\text{(a)}}$Institute for Theoretical Physics, J.W. Goethe University,
Max-von-Laue-Str.\ 1, D--60438 Frankfurt am Main, Germany}
\address{$^{\text{(b)}}$Dip.~di Fisica dell'Universit\`a di Ferrara and INFN Sez.~di
Ferrara, Via Saragat 1, I-44100 Ferrara, Italy}

\section{Introduction}

In the experimental work of Ref. \cite{Litvinov:2008rk} non-exponential decays
of Hydrogen-like ions $^{140}$Pr and $^{142}$Pm have been observed. Denoting
$N(t)$ as the number of unstable particles at the instant $t$, one finds that
$dN/dt$ does $\emph{not}$ follow a simple exponential law of the form
$e^{-\lambda t}$, but shows superimposed oscillations fitted by the following
formula:
\begin{equation}
\frac{dN_{dec}}{dt}=-\frac{dN}{dt}\propto e^{-\lambda t}(1+a\cos(\omega
t+\phi))\text{ ,} \label{fitgsi}%
\end{equation}
where $dN_{dec}/dt$ represents the number of decay per time (see Fig. 3-5 of
Ref. \cite{Litvinov:2008rk}). These unexpected oscillations on top of the
exponential decay are known as the `GSI anomaly'. The possible explanations of
the observed experimental data by invoking neutrino oscillations, neutrino
spin precession and quantum beats seem not to be satisfactory, see Refs.
\cite{Cohen:2008qb,gal,Merle:2010qq,Merle:2009re,Wu:2010ke,giunti} and refs. therein.

In the framework of non-relativistic quantum mechanics it is indeed known that
the exponential decay law is only an approximation, a very a good one, which
however holds at late times after the `preparation' of the unstable system,
see Refs.
\cite{1978RPPh...41..587F,peres,1996IJMPB..10..247N,2002quant.ph..2127F} and
refs. therein. By observing the system soon after its preparation, deviations
from the exponential decay law occur, which for instance lead to the so called
Quantum Zeno effect \cite{1977JMP....18..756M}. Similar properties have been
found also in the framework of a genuinely relativistic quantum field
theoretical approach \cite{Giacosa:2010br}. Such deviations are however very
difficult to observe because they usually occur on very short time scales, of
the order of $10^{-15}$ s for electromagnetic atomic decays
\cite{1998PhLA..241..139F} and even shorter for strong decays
\cite{Giacosa:2010br}. A renewed interest in this topic appeared when recent
cold atom experiments allowed to clearly observe for the first time deviations
from the exponential decay law of unstable systems (via tunneling of atoms out
of a trap) \cite{raizen}. Also the Quantum Zeno effect has found its
experimental evidence \cite{raizen1} and there is now a growing area of
research related to the so called Quantum Zeno Dynamics
\cite{2008JPhA...41W3001F}.

It is tempting to interpret the deviations from the exponential law measured
at GSI as a pure quantum mechanical effect which has never been observed in
the past due to the technical difficulties in doing these kind of experiments.
The GSI setup is indeed unique because of its capability of \textquotedblleft
creating\textquotedblright\ the unstable states at $t=0$, H-like ions in their
case, and to follow their temporal evolution for the first few tens of seconds
by single-particle decay spectroscopy. Notice also that they ``create'' only a
few unstable ions for each run of the experiment and collect some thousand
measurements to get a good statistics. On the other hand, in a measurement of
the decay probability of the same species by using a ``chunk'' of unstable
ions, the effect should be less evident.

In this work we aim to show that it is possible to understand the oscillations
such as the ones measured in the GSI experiment by using only quantum
mechanics. In fact, oscillations which are qualitatively very similar to the
measured ones are obtained as soon as one goes beyond the simple Breit-Wigner
\cite{breit} form for the mass distribution of the unstable resonance. Indeed,
in a simple and solvable model for the decay developed in Ref. \cite{winter},
oscillations are found at short times and long times after the preparation of
the unstable state and represent the transitions from the initial quadratic
behavior of the survival probability and the exponential law and then from the
exponential law to the power law.
Also in the framework of electromagnetic atomic decays, a similar phenomenon
has been described in Ref. \cite{1998PhLA..241..139F}.

\section{Decay law of quantum systems: a phenomenological model}

In order to discuss our interpretation of the oscillations seen in the GSI
experiment, we need first to briefly review the basic formulae concerning the
decay law of an unstable state. The decay survival amplitude $a(t)$ and the
survival probability $p(t)$ can be written as%
\begin{equation}
a(t)=\int_{-\infty}^{\infty}\mathrm{dx}d(x)e^{-ixt\text{ }}\text{ ,
}p(t)=\left\vert a(t)\right\vert ^{2}\text{ .}%
\end{equation}
In the Breit-Wigner case one has
\begin{equation}
d(x)\rightarrow d_{BW}(x)=N\frac{\Gamma_{0}}{(x-M)^{2}+\Gamma_{0}^{2}/4}\text{
},
\end{equation}
where $M$ is the mass of the resonance (i.e. its energy in the rest frame),
$\Gamma_{0}$ its decay width and $N=1/2\pi$ is the normalization assuring that
$a(0)=1$ (the state is prepared at the instant $t=0$ with unit probability,
$p(0)=1$). As a consequence, when doing the Fourier transform of the
Breit-Wigner distribution, the integral gets only the contribution of the
simple pole located at $x_{pole}=M-i\Gamma_{0}/2$ leading to
\[
a_{BW}(t)=e^{-iMt}e^{-\Gamma_{0}t/2}\text{ }.
\]
The usual exponential law for the survival probability, i.e. the probability
that the unstable state did $not$ decay at the instant $t$, emerges:
\begin{equation}
p_{BW}(t)=\left\vert a_{BW}(t)\right\vert ^{2}=e^{-\Gamma_{0}t}\text{ }.
\end{equation}

However, the Breit-Wigner distribution used above is only an approximation,
which turned out to be very useful in many practical cases but is nonetheless
for very basic reasons not exact, because it relies on the assumption that the
spectrum is unbounded from below. Moreover, the average quantities
$\left\langle E\right\rangle $ and $\left\langle E^{2}\right\rangle $ are not
defined. It is therefore expected that far away from the peak the mass
distribution $d(x)$ decreases \emph{faster} than $1/x^{2}$.

\begin{figure}[ptb]
\vspace{0.5cm} \begin{centering}
\epsfig{file=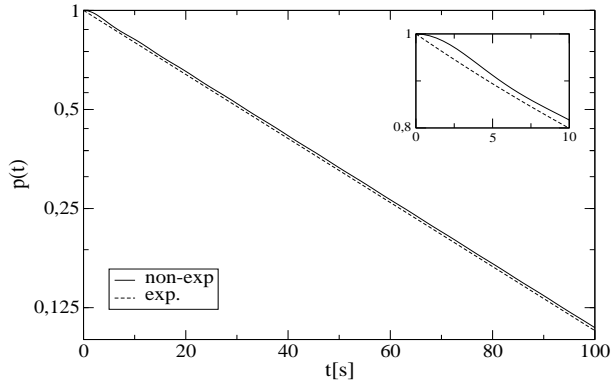,height=8cm,width=5cm,angle=-90}
\caption{Survival probability as a function of time for our modified
Breit-Wigner distribution
(solid line) which shows deviations from the pure exponential decay law (dashed
line). The insert
displays the deviations from the exponential at times close to $t=0$.}
\end{centering}
\end{figure}

In a general framework, the quantity $\Gamma_{0}$ should be replaced by a
function dependent on the energy, $\Gamma_{0}\rightarrow\Gamma(x)$. The simple
inclusion of an energy threshold, $\Gamma(x)=0$ for $x\leq x_{0}$, and
provided that the average energy is defined, is already enough to change the
properties of the system for both short and large times. In both cases the
exponential behavior is not realized: for short times $p(t)$ shows the
quadratic behavior $p(t)=1-\#t^{2},$ and for large times a power-law of the
kind $p(t)\sim t^{-n}$ takes place \cite{1978RPPh...41..587F}. More in
general, when including loops, also the real part of the self-energy
--neglected here-- would play a role, which assures the correct normalization
of the mass distribution (this is a consequence of the K\"{a}llen-Lehman
representation, see Refs. \cite{Achasov:2004uq,Giacosa:2007bn} and refs.
therein). Concerning the analytical structure of the distribution, as it has
been discussed in Ref. \cite{2002quant.ph..2127F} within the Lee model, the
presence of the branch-cut and the specific details of the form factor
determine the deviation from the exponential decay law.

The decay under study in Ref. \cite{Litvinov:2008rk} is a weak decay of an ion
with a characteristic decay rate $\Gamma_{0}$ of $1$ min$^{-1}$ and a $Q$
factor of a few MeV. It is therefore clear that in this case $\Gamma/M$ is
very small and it is technically convenient to perform the change in the
integration variable, $x-M=y,$ thus obtaining:
\begin{equation}
a(t)=e^{-iMt}\int_{-\infty}^{\infty}dyN\frac{\Gamma(y)}{y^{2}+\Gamma(y)^{2}%
/4}e^{-iyt\text{ }}\text{ .}%
\end{equation}
One knows that $\Gamma(y)\simeq\Gamma_{0}$ for $y$ not far from the peak in
$y=0$. However, as already remarked before, the distribution far away from the
peak should decrease faster than $1/y^{2}$. In this work we take into account
this fact by modelling the energy dependence of the decay width $\Gamma(y)$ in
a simple phenomenological way:%
\begin{equation}
\Gamma(y)=\Gamma_{0}\theta(y+\Lambda_{1})\theta(\Lambda_{2}-y)\text{ ,}
\label{cuts}%
\end{equation}
where $\Lambda_{1}$ and $\Lambda_{2}$ are positive numbers, which should be
obviously larger than the mean width of the peak $\Gamma_{0}.$ We have thus
cut the distribution on the left and on the right sides of the peak. These
corrections are important because they encode deviations from the Breit-Wigner
behavior for values of the energy far away from the peak, whose physical
origin lies in the microscopic properties of the form factors and/or in the
interaction with the experimental apparatus, see details later on.
Independently on the origin of the cutoff(s), the simple parametrization used
here is sufficient to show the underlying mechanism and the emergence, as we
will discuss in the following, of oscillations superimposed to the exponential
decay law in a quite general and understandable framework. In the first
section of the Appendix we will also show the results obtained when adopting a
function $\Gamma(y)$ which vanishes smoothly at energies far from the peak.

\bigskip

\begin{figure}[ptb]
\vspace{0.5cm} \begin{centering}
\epsfig{file=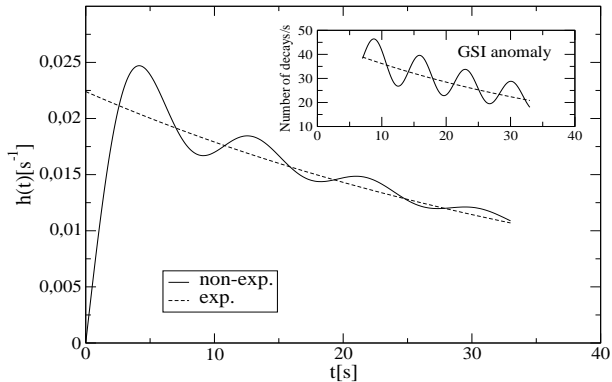,height=8cm,width=5cm,angle=-90}
\caption{$h(t)$ as a function of time for
the non-exponential (solid line) and the exponential (dashed line) cases.
Sizable oscillations are superimposed on the standard exponential
decay for $t \sim 0$. The insert shows a fit that was performed
in \cite{Litvinov:2008rk} on the data of the GSI anomaly. Our model,
having only {\it one free parameter}, can qualitatively reproduce
the observed oscillations. Our curves and the fit have different normalizations.}
\end{centering}
\end{figure}

The survival probability $p(t)$ represents the probability that a state
prepared at $t=0$ did not decay at the instant $t$. Therefore when $N_{0}$
states are present for $t=0$, the number of states changes as $N(t)=N_{0}%
p(t)$. In agreement with the experimental analysis of Ref.
\cite{Litvinov:2008rk}, we are interested in the derivative $dN/dt$, i.e. in
the number of decays over time $dN_{dec}/dt$, which is given by:%
\begin{equation}
\frac{dN_{dec}}{dt}=-\frac{dN}{dt}=N_{0}h(t)\text{ with }h(t)=-\frac
{dp(t)}{dt}\text{ .} \label{h}%
\end{equation}
In the previous equation the function $h(t)$ has been introduced, whose
physical interpretation is straightforward: $h(t)dt$ represents the
probability that one unstable state decays within $t$ and $t+dt.$ It is also
denoted as `decay rate' or `decay probability density'.

Let us show now the results we obtain by using our simple model. We focus here
on the case of $^{142}$Pm of Ref. \cite{Litvinov:2008rk} for which the total
decay rate, obtained from the fit to the data by using the modified
exponential, is $\Gamma_{0}=0.0224$ s$^{-1}$ ($\sim10^{-17}$ eV). We use this
number as the main time/energy scale in our model. In the present case the
threshold $E_{0}\simeq10^{23}\Gamma_{0}$ is too far to have any physical
effect; moreover, we work with the additional assumption $\Lambda_{1}%
=\Lambda_{2}=\Lambda$, i.e. the peak is symmetrically restricted on both
sides, thus leaving us with only one free parameter. In Fig. 1 we show the
survival probability for the numerical choice $\Lambda=32\Gamma_{0}$: at large
times the standard exponential decay law is correctly obtained and at small
times (see the insert) sizable deviations from the exponential are present. In
Fig. 2 we show the function $h(t)$ defined in Eq. (\ref{h}). In the insert we
have also displayed the curve obtained in Ref. \cite{Litvinov:2008rk} by
fitting the experimental data using Eq. (\ref{fitgsi}). Quite remarkably, we
can qualitatively reproduce the oscillations observed in the experiments. A
part from the normalization which for us is one but for the experimental fit
depends on the number of injected ions, the frequency of the oscillation is
correctly reproduced by fixing the only free parameter $\Lambda=32\Gamma_{0}$.

Two basic differences of our curve w.r.t. the fitting curve of Ref.
\cite{Litvinov:2008rk} are visible: (i) the first peak of our oscillations is
more pronounced than the others and (ii) the amplitude of the oscillations is,
in our case, more suppressed as the time increases. Interestingly, by looking
at the data points of Fig. 5 of Ref. \cite{Litvinov:2008rk} both properties
(i) and (ii) hold: the first two points are more than $\sim2\sigma$ above the
exponential curve and the last few points are almost on top of the exponential
curve. \begin{figure}[ptb]
\vspace{0.5cm} \begin{centering}
\epsfig{file=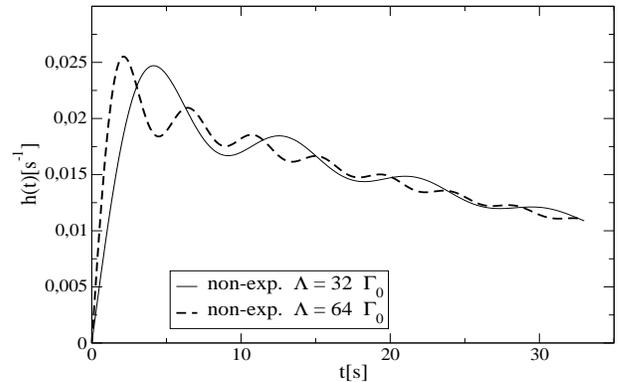,height=8cm,width=5cm,angle=-90}
\caption{$h(t)$ for two different choices of the cutoff. As the cutoff increases,
the frequency also increases and the amplitude of the oscillation decreases.}
\end{centering}
\end{figure}Of course, one would actually need the data to perform a detailed
fit within our model and check the value of the $\chi^{2}$, nevertheless, the
very fact that the superimposed oscillations naturally emerge is quite
exciting and can help to understand the GSI anomaly as a genuine quantum effect.

Moreover, as our Fig. 2 clearly shows, we can make a simple prediction
concerning the short time behavior: the function $dN_{dec}/dt$ is expected to
be sizably smaller for $t\lesssim5$ s and should tend to zero for
$t\rightarrow0$. This is a specific feature of our interpretation of the GSI
anomaly and it is related to the well known fact that decay probability
approaches $t=0$ quadratically as predicted by Quantum Mechanics. If the
experiment at GSI could measure a few points below $10$ s, our interpretation
could be easily rejected or approved.

Let us comment briefly on the dependence of the oscillation frequency $\omega$
and the amplitude on the specific unstable nucleus. In Ref.
\cite{Litvinov:2008rk}, there is an indication that $\omega$ and the amplitude
are anti-correlated (only two ions are however examined). This feature is
present in our model. In Fig. 3 we show our $h(t)$ at fixed $\Gamma_{0}$ and
with cutoffs of $32\Gamma_{0}$ and $64\Gamma_{0}$. By increasing the cutoff,
$\omega$ increases and the amplitude (at fixed time) decreases. We will
discuss the specific case of the $^{140}$Pr ions in the last section of the Appendix.

\section{Interpretation of the GSI anomaly}

We now comment on the physical interpretation of these oscillations. Similar
oscillations of the decay probability have also been found in Ref.
\cite{2002quant.ph..2127F}. The authors use a nonrelativistic quantum field
theory, by employing Lee Hamiltonians, to show that the complicated analytical
structure of the propagator in the complex energy plane can be captured by a
simple two-pole model: the first pole corresponds to the usual one which
defines the mass and the decay rate of the state, the second pole corresponds
instead to the leading order contribution of the branch-cut to the decay
amplitude. Interestingly, in their \textquotedblleft small coupling
limit\textquotedblright\ they find an exponential decay modulated by
oscillation which are qualitatively similar to the ones here presented. A
simple physical interpretation emerges: this peculiar behavior of the decay
probability interpolates between the pure exponential decay which occurs in
presence of a large bandwidth continuum into which the unstable state can
decay, and a pure oscillating probability which occurs in a system of two
discrete levels where Rabi oscillations are obtained.

In our model, the cutoff regulates the spread of the energy of the continuum
of states to which the unstable state is coupled. The oscillating behavior is
clearly visible only if the cutoff is smaller than -say- $100\Gamma_{0}$: the
unstable state and the continuum are coupled only in a (relatively) small
window of energy and thus, as in the two-pole model, the system shows
`Rabi-like oscillations' in the decay probability. Besides the similarities,
we also point out an important difference w.r.t. Ref.
\cite{2002quant.ph..2127F}. While in our case the function $h(t)$ is always
positive, the oscillations found in Ref. \cite{2002quant.ph..2127F} allow also
for zero and negative values of $h(t)$; in that case the Rabi-like
oscillations are much stronger.


\section{Discussion and conclusions}

We now discuss the physical origin of the cutoff that we use to describe the
data. A first possibility is that our cutoff emerges as a \textquotedblleft
natural cutoff\textquotedblright\ from the calculation of the evolution of the
initial unstable state by use of Nuclear Physics models of the nucleus and its
weak and electromagnetic interactions with the orbiting electron (e.g. Ref.
\cite{Bambynek:1977zz}). For instance, in the case of the electromagnetic
atomic decay studied in Ref. \cite{1998PhLA..241..139F}, such a natural cutoff
appears in the wave functions of the electrons and scales as the fine
structure constant $\Lambda\propto\alpha$. The microscopic calculation of form
factors is technically very difficult, we suspect however that it is unlikely
that the strong and the electromagnetic interactions, which would be
responsible for the form factor for the decay here analyzed, could explain the
existence of a cutoff at the energy scale of $\sim10^{-16}$ eV needed to
describe the GSI result. A second, more promising, possibility is that our
cutoff emerges from the interaction of the unstable system and the
experimental apparatus. It has been indeed already stressed in Refs.
\cite{1978RPPh...41..587F,degasperis,koshino} that the measurement itself is
an interaction between two physical systems and a cutoff appears which is
connected with the finite response time of the experimental apparatus.
(Indeed, in Refs. \cite{1978RPPh...41..587F,degasperis} the very same Eq.
(\ref{cuts}) has been introduced as an effect of the interaction with the
measuring apparatus.) During this temporal window the correlation between the
unstable state and the decay products is not destroyed by the measurement. The
response time is usually very small (and the corresponding cutoff large) for
standard decay experiments. In the GSI experiment, the response time $\delta
t$ should be related, for instance, to the time precision between subsequent
observations, which was of the order of one fifth of a second, and/or to the
time needed to cool the daughter nucleus before its detection in the storage
ring, which was of the order of one second. Remarkably, the inverse of this
temporal scale is very close to the cutoff scale $\Lambda$ used to describe
the data ($\Lambda\sim1/\delta t$). Moreover, in the framework of this
interpretation it immediately follows that a repetition of the experiment with
an improved time precision (or in general a smaller response time) would
correspond to a higher cutoff $\Lambda.$ For instance, by increasing the time
resolution of a factor 2 implies also a cutoff which is larger of a factor 2:
as shown in Fig. 3, in this case the amplitude and the period of the
oscillations decrease. The very same argumentation can be used to explain why
in the experiment at the Berkeley Lab \cite{vetter} no oscillations have been
observed, see also the detailed discussion of this point in Ref.
\cite{Giacosa:2012yd}. In the Berkeley experiment the $^{142}$Pm nuclei are
embedded in a lattice and are not ionized. When the innermost K-shell electron
is captured by the nucleus, an electron from the outer levels jumps to the
ground state and a photon is immediately emitted, which is then absorbed by
the environment and/or the detector after a very short time delay. In this
case, the time-energy uncertainty relation implies that the cutoff is much
larger than the decay width, $\Lambda_{ec}\gg\Gamma_{0}$. As visible in Fig.
3, when the cutoff increases the period of the oscillation decreases. Thus, in
the case of the Berkeley experiment the oscillations are extremely suppressed
and cannot be detected. Our approach, together with the assumption of a cutoff
induced by the measurement, can explain in a rather natural way the absence of
oscillations in the experimental setup of Ref. \cite{vetter}. Furthermore, it
is also possible to study the other decay channel of $^{142}$Pm, the
$\beta^{+}$ process, and to explain why also in this case no oscillations
should be observed (see section A.2 in the Appendix).

Another subtle issue emerges when trying to interpret the results of the GSI
by using the standard \textquotedblleft projection postulate\textquotedblright%
\ of Quantum Mechanics: in an ideal measurement the collapse of the wave
function, and thus the measurement itself, occurs instantaneously as soon as
the wave function of the unstable system interacts/overlaps with the
measurement apparatus. In this sense the measurement effectively
\textquotedblleft resets the clock\textquotedblright\ every time the unstable
system is found to be undecayed and what could be measured, at the storage
ring of the GSI for instance, is just an exponential decay, which is slowed
down (quantum Zeno effect) or accelerated (quantum Anti-Zeno effect) depending
on the details of the unstable state and on the time interval of the pulsed
measurements. On the other hand, the measurement at the GSI experiment is
quite peculiar: the system \textquotedblleft sees\textquotedblright\ (in the
frequency spectrum) the ions averagely every $\sim200$ ms and not at every
passage through the mass spectrometer, which occur every $0.5\mu$s. Moreover,
as discussed before, the appearance in the frequency spectrum of the daughter
ions is delayed by $900$ and $1400$ ms (for $^{140}$Pr and $^{142}$Pm
respectively) needed for the cooling. This means that there is a period of
about $1$ s during which the experimental apparatus does not \textquotedblleft
see\textquotedblright\ neither the parent ion nor the daughter ion. This
measurement clearly is not an ideal measurement: a detailed modelling of the
innovative and unique measuring procedure used at GSI on the line of the
theory of measurement described in Refs. \cite{koshino,degasperis,kurizki}
deserves further investigation. Notice also that, within our interpretation,
the experiment at GSI would be a precious experimental apparatus for the study
of the fundamental open questions related to the process of measurement in
Quantum Mechanics.

In conclusions, we have studied in a general framework a typical behavior of
unstable particles: exponential decay law with superimposed oscillations. We
have shown that this behavior is quite common as soon as the Breit-Wigner
distribution of the unstable state is left and (even very simple) form-factors
are taken into account, which suppress the Breit-Wigner distribution far away
from the peak. Using a cut-off model, we have shown that we can reproduce the
qualitative behavior of the oscillations seen in the GSI experiment.
Obviously, the adopted modification represents a first attempt to investigate
how the oscillations in the decay law emerge. Future work in this direction
should go beyond the simple cutoff. For instance, the detailed modelling of
the measurement should also naturally provide how the energy distribution
deviates from the Breit-Wigner form. However, independently on the details of
the deviations from the Breit-Wigner form, if our interpretation is correct,
we predict that, for times smaller than $5$ s (see Fig. 2), the number of
decays per seconds rapidly drops to zero due to a fundamental property of
quantum systems: the quadratic behavior of the survival probability at short
times after the preparation of the system. Moreover, we also predict that the
first oscillation is more pronounced than the others.

Finally, we would like to mention that possible indications of the presence of
oscillations superimposed to the exponential decay were found also in two
other unstable, but utterly different, quantum systems: the tunneling of cold
atoms out of a trap \cite{raizen} and the decay of $^{32}$Si
\cite{1986E&PSL..78..168A}. In particular, in the latter case, the unstable
nuclei $^{32}$Si have a very large half-life of about 170 yr, but show
superimposed oscillations of about 1 yr. We speculate that also these
superimposed oscillations represent a manifestation of the same fundamental
phenomenon of Quantum Mechanics.

\bigskip

\textbf{Acknowledgments:} The authors thank G. Torrieri and H. Warringa for
pointing out Refs. [1,19] and A. Merle for very useful comments. G.P.
acknowledges financial support from the Italian Ministry of Research through
the program \textquotedblleft Rita Levi Montalcini\textquotedblright.

\bigskip

\appendix

\section{}


In this appendix we discuss in more details three subjects: the existence of
oscillations for smooth form factors, the inclusion of two decay channels and
finally the case of the decay of the $H$-like $^{140}$Pr ions also measured at GSI.

\subsection{Oscillations upon variation of form factors}

The existence of oscillations does not depend on the precise form of the
cutoff. In Ref. \cite{Giacosa:2012yd} the cases in which only one end is open
have been tested. Here we study the oscillations by introducing a Fermi-like
cutoff function of the form:
\begin{align}
a(t)  &  =N\int_{-\infty}^{\infty}\frac{\Gamma(y)}{y^{2}+\Gamma^{2}%
(y)/4}\text{ }\\
\Gamma(y)  &  =\frac{\Gamma_{0}}{1+e^{-\alpha^{2}(y^{2}-\Lambda^{2})}}\text{
,}%
\end{align}
where the constant $\alpha$ controls the steepness of the fall-off for
$y\simeq\pm\Lambda$. In Fig. 4, we show the results obtained for $h(t)$ for
three different choices of the parameter $\alpha$. Clearly the smaller the
value of $\alpha$ the smaller are the oscillations superimposed to the
exponential decay. This fact has a natural explanation by reminding that the
oscillations are present only if the bandwidth of the continuum of states into
which the unstable state decays is narrow, as explained before. We notice on
the other hand that it is possible to obtain very large oscillations (with
$h(t)$ reaching even negative values) within the two pole model proposed in
\cite{2002quant.ph..2127F}. More in general, one can extend the present study
by investigating more complicated forms of the function $\Gamma(y).$ This
represents an interesting outlook for future work.

\bigskip

\begin{figure}[ptb]
\vspace{0.6cm} \begin{centering}
\epsfig{file=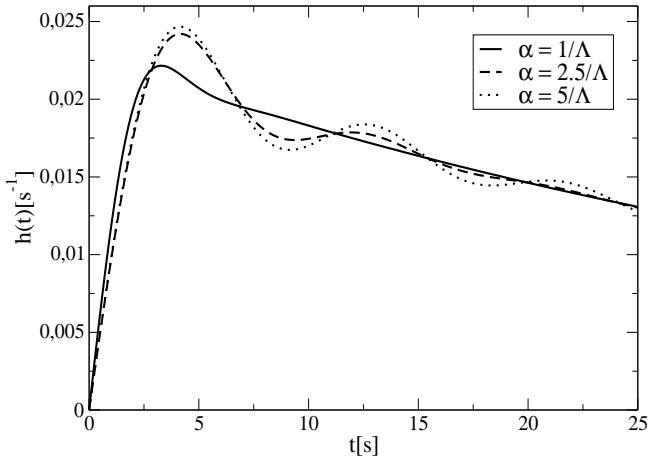,height=8.5cm,width=6cm,angle=-90}
\caption{$h(t)$ is shown in the case of a smooth cutoff function (Fermi-like). The smaller
the value of $\alpha$ the smaller the oscillations are.}
\end{centering}
\end{figure}

\subsection{Two channels}

We turn now to the non-exponential decay when two decay channels are present.
To this end we use the formalism developed in Ref. \cite{duecan}. The total
energy-dependent decay width reads $\Gamma(y)=\Gamma_{1}(y)+\Gamma_{2}(y)$,
where $\Gamma_{i}(y)$ represent the energy-dependent decay width in the $i$-th
channel. The survival probability amplitude $a(t)$ can be decomposed as the
sum of two terms:
\begin{equation}
a(t)=N\int_{-\infty}^{\infty}\frac{\Gamma(y)}{y^{2}+\Gamma^{2}(y)/4}%
=a_{1}(t)+a_{1}(t)
\end{equation}
whereas $N$ assures that $a(0)=1$ and
\begin{align}
a_{1}(t)  &  =N\int_{-\infty}^{\infty}\frac{\Gamma_{1}(y)}{y^{2}+\Gamma
^{2}(y)/4}\text{ ,}\\
\text{ }a_{2}(t)  &  =N\int_{-\infty}^{\infty}\frac{\Gamma_{2}(y)}%
{y^{2}+\Gamma^{2}(y)/4}\text{ .}%
\end{align}
The decay probability densities per each channel $h_{1}(t)$ and $h_{2}(t)$
read \cite{duecan}:
\begin{equation}
h_{1}(t)=-A_{1}^{\prime}(t)-A_{mix}^{\prime}(t)\text{ , }h_{2}(t)=-A_{2}%
^{\prime}(t)-A_{mix}^{\prime}(t)\text{ } \label{h12}%
\end{equation}
with
\begin{align}
A_{1}(t)  &  =\left\vert a_{1}(t)\right\vert ^{2}\text{ , }A_{2}(t)=\left\vert
a_{2}(t)\right\vert ^{2}\text{ , }\\
A_{mix}(t)  &  =\operatorname{Re}\left[  a_{1}(t)a_{2}^{\ast}(t)\right]
\text{ .}%
\end{align}
In the case of the $H$-like ion $^{142}$Pm, one has that $\Gamma_{ec}%
=\Gamma_{1}(y=0)=0.2\Gamma_{0}$ corresponds to the electron-capture decay
$M\rightarrow D+\nu_{E}$ , and that $\Gamma_{\beta^{+}}=\Gamma_{2}%
(y=0)=0.8\Gamma_{0}$ corresponds to the $\beta^{+}$ decay $M\rightarrow
D^{\prime}+e^{+}+\nu_{E}$. Moreover, following the discussion in the text, we
assign a different cutoff per each channel: $\Gamma_{1}(y)=\Gamma_{ec}%
\theta(y^{2}-\Lambda_{ec}^{2})$ and $\Gamma_{2}(y)=\Gamma_{\beta^{+}}%
\theta(y^{2}-\Lambda_{\beta^{+}}^{2}).$

As described in Sec. II, for the ion $^{142}$Pm a cutoff $\Lambda
_{ec}=32\Gamma_{0}$ reproduces the correct oscillation frequency. The origin
of $\Lambda_{ec}$ can be traced back either to a (quite unnatural) microscopic
form factor or to (more plausible) measurement apparatus effect via the
time-energy uncertainty relation, $\Lambda_{ec}\sim\frac{1}{\delta t}%
\sim10^{-15}$ eV. In the latter case, we can easily estimate the cutoff
$\Lambda_{\beta^{+}}$ in the $\beta^{+}$-channel. The positron is emitted with
an energy within $0$-$4$ MeV. For our estimate let us consider a positron with
$2$ MeV, which corresponds to a speed of $v_{e^{+}}\simeq0.96c.$ Thus, the
positron is absorbed very fast by the environment. Assuming that it travels 1
m, we get $\delta t\simeq10^{-9}$ s. In turn, the cutoff in this channel reads
$\Lambda_{\beta^{+}}\sim10^{-7}$ eV, which corresponds to roughly
$10^{9}\Gamma_{0}.$

In Fig. 5-6 we show the functions $h_{1}(t)$ and $h_{2}(t).$ The function
$h_{1}(t)$, which describes the decay probability density in the
electron-capture decay channel, shows a behavior which is very similar to the
one of Fig. 3 in which the $\beta^{+}$ decay channel was not considered. On
the contrary, the function $h_{2}(t)$, which describes the decay probability
density in the $\beta^{+}$ decay channel, is not distinguishable from an
exponential decay for times larger than $\sim5$s. Notice that, for our choice
of parameters, the first peak is present also in the $\beta^{+}$ channel due
to the mixing of the two channels implied by formulae (\ref{h12}). On the
other hand, by fixing also $\Lambda_{ec}$ to be large (as in the case of the
Berkeley experiment), both $h_{1}$ and $h_{2}$ would be basically pure
exponential laws.

\bigskip

\begin{figure}[ptb]
\vspace{0.5cm} \begin{centering}
\epsfig{file=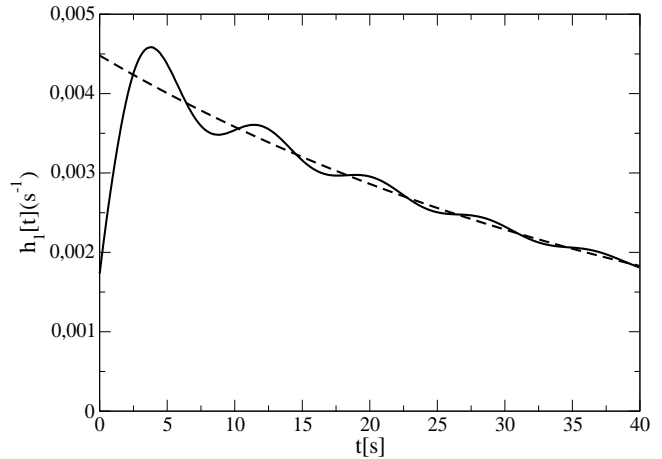,height=8.5cm,width=6cm,angle=-90}
\caption{Decay rate $h_1$ for the electron capture channel. The dashed line corresponds to the exponential decay.}
\end{centering}
\end{figure}

\bigskip

\begin{figure}[ptb]
\vspace{0.8cm} \begin{centering}
\epsfig{file=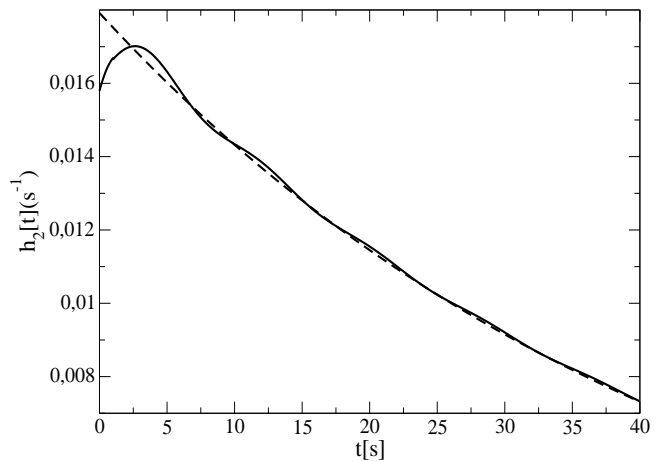,height=8.5cm,width=6cm,angle=-90}
\caption{Decay rate $h_2$ for the $\beta^+$ channel. The dashed line corresponds to the exponential decay.}
\end{centering}
\end{figure}

\subsection{Changing the $H$-like ion}

As a last step, we briefly comment on the decay of the other ion, $^{140}$Pr,
measured in Ref. \cite{Litvinov:2008rk}. The modulation frequency for $^{140}%
$Pr is slightly higher than the one of $^{142}$Pm ($0.890$ s$^{-1}$ and
$0.885$ s$^{-1}$ respectively). In our interpretation, if the cutoff
originates from the experimental apparatus and in particular from its response
time (temporal resolution and cooling time as discussed before), we expect a
mild dependence from the mass number $A$ and charge $Z$ of the nucleus. The
cooling times for the decay products of $^{140}$Pr and $^{142}$Pm are
respectively $900$ ms and $1400$ ms (corresponding to their different recoil
energy). By assuming that the temporal resolution for the two ions is the
same, one would expect that the modulation frequency of $^{140}$Pr is larger
than the one of $^{142}$Pm as the data seem to indicate. Again, at this level,
it is complicated to provide quantitative estimates and moreover it would be
mandatory to experimentally check the dependence of the modulation frequency
from the nucleus by using other species and with more accurate precision.

\bigskip


\end{document}